\documentclass{goose-article}

\hypersetup{pdfauthor={%
  T.W.J. de Geus%
}}
\header{%
  T.W.J.~de~Geus, R.H.J.~Peerlings, M.G.D.~Geers\\%
  Engineering Fracture Mechanics, 2015, 147:318-330, %
  \doi{10.1016/j.engfracmech.2015.04.010}%
}

\title{%
  Microstructural modeling of ductile fracture initiation in multi-phase materials%
}
\author[1,2]{T.W.J.~de~Geus$^*$}
\author[1]{R.H.J.~Peerlings}
\author[1]{M.G.D.~Geers}
\affil[1]{%
  Department of Mechanical Engineering, Eindhoven University of Technology, Eindhoven, The Netherlands%
}
\affil[2]{%
  Materials innovation institute (M2i), Delft, The Netherlands%
}
\contact{%
  $^*$Corresponding author:
  \href{mailto:t.w.j.d.geus@tue.nl}{t.w.j.d.geus@tue.nl} %
  \hspace{1mm}--\hspace{1mm} %
  \href{mailto:tom@geus.me}{tom@geus.me} %
  \hspace{1mm}--\hspace{1mm} %
  \href{http://www.geus.me}{www.geus.me}%
}

\begin{document}

\maketitle

\begin{abstract}
The precise mechanisms underlying the failure of multi-phase materials may be strongly dependent on the material's microstructural morphology. Micromechanical modeling has provided much insight into this dependence, but uncertainties remain about crucial modeling assumptions. This paper assesses the influence of different grain shapes, damage indicators, and stress states using a structured numerical model. A distinct spatial arrangement of phases around fracture incidents is found, consisting of hard regions in the tensile direction interrupted by soft regions in the directions of shear. These key features are only mildly sensitive to the studied variations.
\end{abstract}


\section{Introduction}

Multi-phase materials, such as dual phase steel, metal matrix composites, etc., are frequently used in engineering applications. These materials often compromise strength with ductility. This favorable combination of properties is achieved by combining two or more phases at the level of the microstructure, for example hard (yet brittle) particles embedded in a soft (ductile) matrix. Although the macroscopic elasto-plastic and hardening behavior may be reasonably well predicted for a given microstructure \citep{Choi2009, Sun2009, Deng2006, Heinrich2012, Povirk1995}, many uncertainties remain about the dominant failure mechanism(s). Experimental observations based on fractography, in-situ electron scanning microscopy and tomography suggest that failure often occurs by ductile fracture of the, generally relatively soft, matrix phase \citep{Kadkhodapour2011, Avramovic-Cingara2009a, Ahmad2000, Tasan2010, Williams2010}. However, also different mechanisms are observed in dual-phase steel \citep{Ahmad2000, Avramovic-Cingara2009, Avramovic-Cingara2009a, Uthaisangsuk2008}, and in metal-ceramic composites in particular with a comparatively hard matrix phase \citep{Maire2007, Williams2010}.

Several numerical studies have been performed aiming to unravel the complexity of the fracture mechanisms. These models often use a relatively simple representation of the material in which the different phases are considered elasto-plastic, whereby fracture is associated with large local plastic deformation, see \citep{Choi2009, Choi2009a, Sun2009, Sun2009a, Kumar2006, Povirk1995} and others. For example, \citet{Choi2009a} reported that lower levels of damage occur when the hard phase is distributed more homogeneously. Only few studies have performed a systematic analysis of the effect of the local phase distribution on failure. \citet{Kumar2006} generated statistically representative microstructures from which a critical configuration is identified. This so-called ``hot-spot'' consists of a soft region neighbored on both sides by regions of the hard phase. It is often recognized that the local incompatibility triggers both a high hydrostatic stress and high plastic deformation \citep{Mortensen2010}. By combining a large number of different microstructures a similar observation was made by De Geus et al.~\citep{DeGeus2015a}, who identified the average phase distribution around the initiation of fracture. In addition to the observations by \citet{Kumar2006} it was found the band of hard phase in the tensile direction is interrupted by a band of soft phase in the direction of maximum shear.

From a modeling point of view different approaches are used to incorporate and/or study the micromechanics of a two-phase material. Unit cell models have been used to study the basic micromechanical response, including fracture initiation mechanisms \citep{Argon1975, Beremin1981, Bao1991, Tvergaard1990}. To accommodate the geometrical complexity of two- or multi-phase materials, models are needed that include a large number of particles/grains. Models that are based on a real microstructure with all geometrical details, however, suffer from a large number degrees-of-freedom. Furthermore, it is difficult to apply systematic variations in terms of composition and morphology, without changing other parameters at the same time, to unravel their influence on fracture initiation. Therefore, structured models consisting of square elements are frequently used \citep[e.g.][]{Kumar2006, DeGeus2015a}. Moreover, the numerical complexity is often reduced by using simplified damage indicators \citep[e.g.][]{Choi2009, Choi2009a, Sun2009, Sun2009a, Kumar2006, Povirk1995, DeGeus2015a}. It is not trivial to assess how the resulting conclusions are influenced by the approximations made therein.

In particular, in our earlier work \citep{DeGeus2015a}, a microstructure of square equi-sized grains was employed in combination with a simple indicator for fracture initiation, which was based on the well known fact that ductile fracture takes places when a combination exists of a high hydrostatic tensile stress and high plastic deformation. The particular interest was to study which characteristic features in the two-phase microstructure give rise to such conditions. The main finding was that initiation of fracture is strongly governed by the local arrangement of the two phases. A critical arrangement was identified by calculating the average phase distribution around the critical site. It remains, however, questionable to what degree the main conclusions depend on the aforementioned assumptions. The current contribution aims to remedy this concern, by a critical assessment of the effect of the assumptions on the critical phase distribution around fracture initiation. For this purpose, the following analysis steps are made:
\begin{enumerate}
  \item The basic, Rice \& Tracey-like damage indicator is replaced by a more involved Johnson-Cook damage indicator.
  \item The square cells used to represent the individual phases are compared to hexagonal cells, in which (in contrast to the squares) the phases are never connected by a single point.
  \item The applied pure-shear deformation is extended with a volumetric contribution to consider different strain paths resulting in different stress states, which remain proportional throughout the deformation history.
\end{enumerate}
Like in \citep{DeGeus2015a}, this study is limited to the initiation of ductile fracture in the matrix phase, initiation of fracture in the hard phase and in the interface between the hard and the soft phase are not considered. Furthermore, fracture propagation is not considered.

This paper is structured as follows: the microstructural model, including a summary of the main conclusions in \citep{DeGeus2015a}, is discussed in section~\ref{sec:model}. The influence of the respective assumptions are separately discussed in sections~\ref{sec:damage}--\ref{sec:triax}, followed by a discussion and a summary of the conclusions in section~\ref{sec:discussion}.

\section*{Nomenclature}

\begin{tabular}{llll}
$\bm{A}$                                      & second order tensor
&
$\langle a \rangle$                           & ensemble average
\\
$\mathbb{A}$                                  & fourth order tensor
&
$\bar{a}$                                     & volume average
\\
$\mathbb{C} \, = \bm{A} \otimes \bm{B}$       & dyadic tensor product
&
$\lfloor a \rfloor = \tfrac{1}{2} ( a + |a|)$ & positive part of $a$
\\
$\bm{C} = \bm{A} \cdot \bm{B}$                & single tensor contraction
\\
$c \;\, = \bm{A} : \bm{B} = A_{ij} B_{ji}$    & double tensor contraction
\\
\end{tabular}

\section{Reference model and summary of earlier results}
\label{sec:model}

This section describes the model and summarizes the main conclusions reported in \citep{DeGeus2015a}. Several parts of the model are slightly modified, whereby the most important difference is the adopted three-dimensional discretization to allow for more general stress states. Furthermore, the composition of the microstructure is resolved in a weak sense, only allowing fluctuations in the individual microstructural volume elements in the ensemble, as discussed below. The presented results are all generated with the model presented here.

\subsection{Microstructure}

\begin{figure}[htp]
  \centering
  \includegraphics[width=0.9\linewidth]{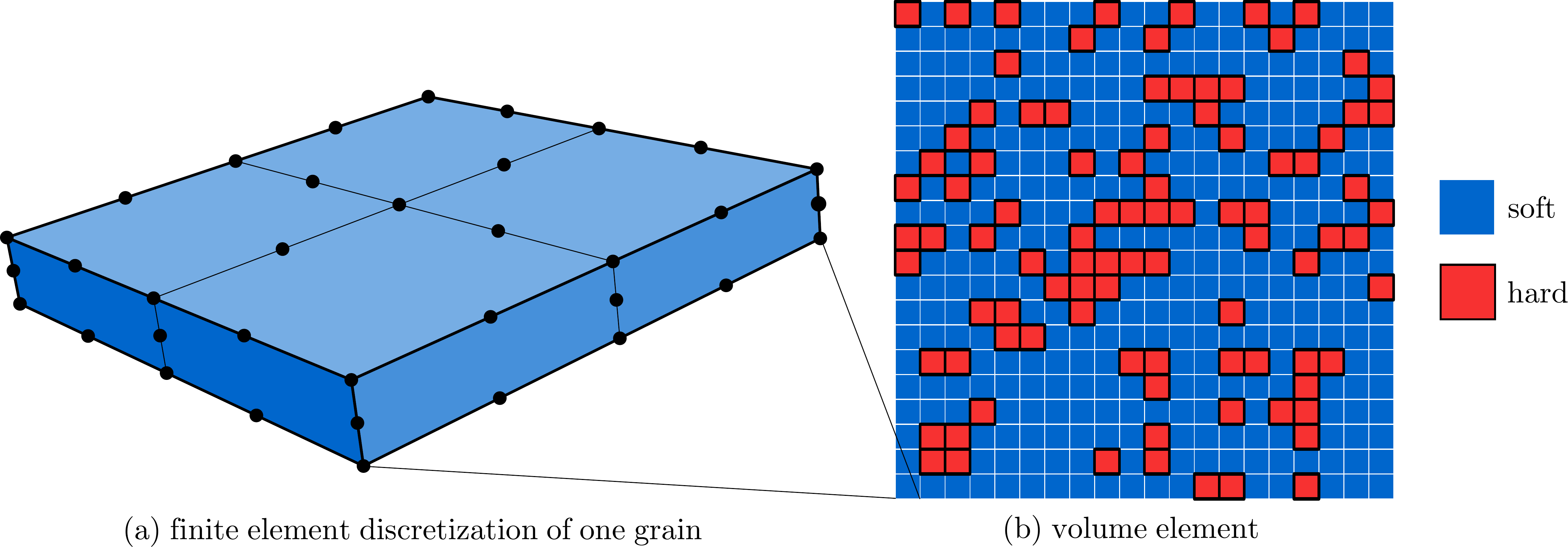}
  \caption{One typical volume element out of the ensemble of $400$ random volume elements in Figure~(b). The finite element discretization of the individual grains is also shown in Figure~(a).}
  \label{fig:typical}
\end{figure}

A two-dimensional microstructure is used that consists of two distinct phases, a comparatively hard phase embedded in a soft phase. An ensemble of 400 randomly generated volume elements is considered, which are assumed periodic to minimize boundary effects. Each of these volume elements comprises $20 \times 20$ equi-sized square cells, representing for example grains in a polycrystal. The influence of the shape of these cells is investigated in Section~\ref{sec:shape}. These grains are randomly assigned the properties of either the hard or the soft phase according to a probability equal to the target volume fraction of hard phase, $\varphi^\mathrm{hard} = 0.25$. This implies that the ensemble averaged volume fraction of hard phase $\langle \varphi^\mathrm{hard} \rangle = 0.25$, while the hard phase volume fraction of individual volume elements ranges between $0.19$ and $0.31$. A typical volume element is shown in Figure~\ref{fig:typical}, where the soft phase is blue and the hard phase is red (with a black outline).

The finite element method is used to calculate the mechanical response of each volume element. In order to apply arbitrary stress states, the two-dimensional microstructure is expanded in thickness direction, and discretized using three-dimensional tri-quadratic cubic finite elements. For each grain one finite element is used in the out-of-plane direction and $2 \times 2$ finite elements are used in the two in-plane directions of Figure~\ref{fig:typical}. Numerical integration is done using eight Gauss-points per finite element.

Given the idealized microstructures only grain-averaged stress and strain measures are computed. To this end, all output tensor components and scalars are volume averaged over the 32 Gauss points in the four finite elements in each grain. It has been verified that this discretization is sufficiently accurate for our purpose, i.e.\ grain averaged quantities do not change significantly upon mesh refinement. Specifically, the maximum local relative error in terms of the used physical quantities is 1\% with respect to a reference discretization of $10 \times 10$ tri-quadratic cubic finite elements per grain.

\subsection{Constitutive model}

Both phases are assumed isotropic elasto-plastic and are modeled using the finite strain model due to \citet{Simo1992a}. This model relies on the multiplicative split of the deformation gradient tensor into an elastic and a plastic part:
\begin{equation}
	\bm{F} = \bm{F}_\mathrm{e} \cdot \bm{F}_\mathrm{p}
\end{equation}
It is defined in the current (deformed) configuration, where the Kirchhoff stress $\bm{\tau}$ is a linear function of the logarithmic elastic strain as follows:
\begin{equation} \label{eq:model:stress}
\bm{\tau} = \tfrac{1}{2} \mathbb{C} : \ln \bm{b}_\mathrm{e}
\end{equation}
with $\bm{b}_e$ the elastic Finger tensor; the elastic stiffness
\begin{equation} \label{eq:model:stiff}
	\mathbb{C} = K \bm{I} \otimes \bm{I} + 2 G \big( \mathbb{I}^s - \tfrac{1}{3} \bm{I} \otimes \bm{I} )
\end{equation}
in which the bulk modulus $K$ and the shear modulus $G$ depend on the Young's modulus $E$ and the Poisson's ratio $\nu$ in the conventional way. Furthermore $\bm{I}$ is the second order unit tensor, and $\mathbb{I}^s$ is the fourth order symmetric unit tensor.

The plasticity is modeled using a $J_2$ criterion in combination with linear hardening. The elastic domain is thus bounded by the following yield criterion
\begin{equation}
	\phi (\bm{\tau},\varepsilon_\mathrm{p}) =
  \tau_\mathrm{eq} - ( \tau_\mathrm{y0} + H \varepsilon_\mathrm{p} ) \leq 0
\end{equation}
where $\tau_\mathrm{eq}$ is the von Mises equivalent stress and $\varepsilon_\mathrm{p}$ the equivalent plastic strain; the hardening modulus $H$ and initial yield stress $\tau_\mathrm{y0}$ are material parameters. Finally, a standard associative flow rule is used. Details on the implementation in the finite element framework (including the consistent linearization) can be found in the work by \citet{Geers2004}.

The elastic properties of the two phases are assumed identical; they differ only through the plastic response. The following parameters are used:
\begin{equation}
	\frac{  \tau_\mathrm{y0}^\mathrm{hard}}{E} =
  \frac{2 \tau_\mathrm{y0}^\mathrm{soft}}{E} = 12 \cdot 10^{-3}
	\qquad
	\frac{  H^\mathrm{hard}}{E} =
  \frac{2 H^\mathrm{soft}}{E} = 16 \cdot 10^{-3}
	\qquad
	\nu = 0.3
\end{equation}
which are representative for a wide class of steels, for example dual-phase steel \citep{Sun2009, Vajragupta2012}.

\subsection{Damage model}

This study is restricted to the initiation of the ductile fracture in the soft phase; the strong hard phase is assumed not to fracture. In this regime, a damage indicator is used to signal the initiation of fracture. This indicator does not affect the mechanical behavior, i.e.\ the material continuously hardens despite the predicted indicator of damage. A simple damage indicator is used that captures the essential characteristics of ductile fracture. It accounts for the influence of plastic strain $\varepsilon_\mathrm{p}$ and a positive hydrostatic stress $\tau_\mathrm{m}$ in a simple linear combination, i.e.\
\begin{equation} \label{eq:model:RT}
	D = \frac{ \varepsilon_\mathrm{p} \lfloor \tau_\mathrm{m} \rfloor }{ D_\mathrm{c} }
\end{equation}
\citep[cf.][and others]{Rice1969, Gurson1977}. Notice that the above formulation is a simplified version of these models. The critical damage $D_\mathrm{c}$ is introduced to compare this damage indicator to the Johnson-Cook model in Section~\ref{sec:damage}. Both models predict a value of $D = 1$ at the same applied uni-axial tensile strain; this results in $D_\mathrm{c} = 2.75 \cdot 10^{-4}$.

\subsection{Applied deformation}

The periodicity of the microstructure is enforced by means of periodic boundary conditions, which couple the average displacement of the boundaries to the macroscopic deformation gradient tensor $\bar{\bm{F}}$. Initially the microstructure is subjected to a macroscopic pure-shear deformation, defined as follows
\begin{equation} \label{eq:model:F}
	\bar{\bm{F}} = \bar{\bm{F}}_\mathrm{d}
  = \exp \left(   \tfrac{\sqrt{3}}{2} \bar{\varepsilon}_\mathrm{d} \right) \, \vec{e}_x \vec{e}_x
  + \exp \left( - \tfrac{\sqrt{3}}{2} \bar{\varepsilon}_\mathrm{d} \right) \, \vec{e}_y \vec{e}_y
  + \vec{e}_z \vec{e}_z
\end{equation}
where $\bar{\varepsilon}_\mathrm{d}$ is the logarithmic stretch ratio. The microstructure is deformed up to $\bar{\varepsilon}_\mathrm{d} = 0.2$ in $200$ increments. In Section~\ref{sec:triax}, the deformation is extended with a volumetric part to consider different stress states.

\begin{figure}[htp]
  \centering
  \includegraphics[width=0.6\linewidth]{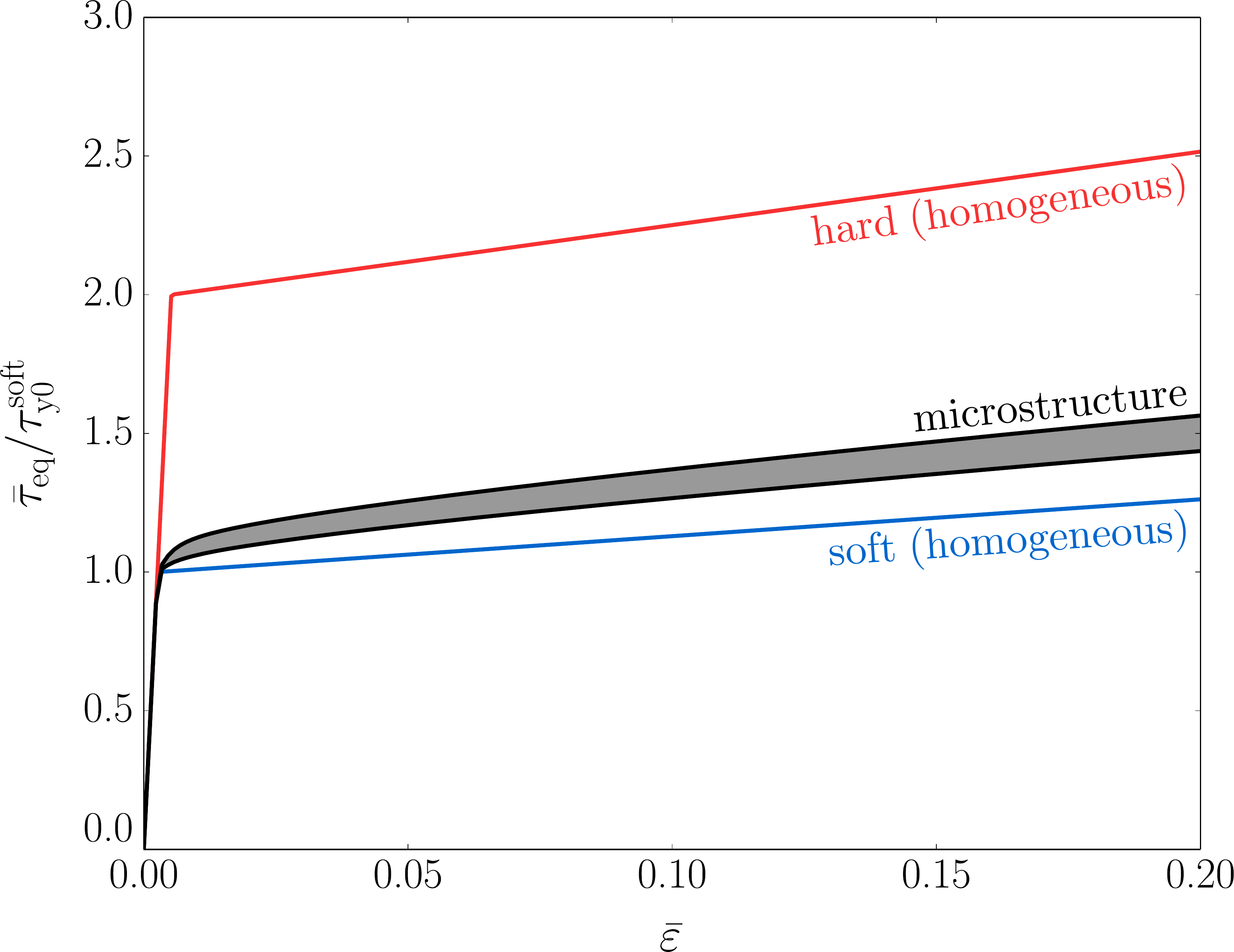}
  \caption{Macroscopic equivalent stress $\bar{\tau}_\mathrm{eq}$ as a function of the applied logarithmic equivalent strain $\bar{\varepsilon}$ in pure shear, for the two homogeneous phases (red and blue) and a band spanning the macroscopic responses of all volume elements in the ensemble of microstructures, with an average volume fraction of hard phase $\langle \varphi^\mathrm{hard} \rangle = 0.25$ (black).}
  \label{fig:stress-strain}
\end{figure}

The band of macroscopic responses of the different volume elements is plotted in Figure~\ref{fig:stress-strain}, with the macroscopic, volume averaged, equivalent stress $\bar{\tau}_\mathrm{eq}$ on the vertical axis and the logarithmic equivalent strain $\bar{\varepsilon}_\mathrm{d}$ on the horizontal axis. The constitutive response of the soft and the hard phase is included using a blue and red line respectively. As observed, the macroscopic response of the microstructure is a non-linear combination of the constitutive response of the individual phases. The scatter that is observed between the individual samples can be directly related to the difference in hard phase volume fraction. The specific arrangement of phases in the volume element has significantly less influence on the macroscopic response. Indeed, when the volume fraction is the same for each volume element hardly any scatter is observed \citep{DeGeus2015a}.

\subsection{Damage hot-spot}

In contrast to the macroscopic response, a strong correlation is found between the local phase distribution and the initiation of fracture. To capture this correlation the so-called ``damage hot-spot'' is calculated, as in \citep{DeGeus2015a}. This analysis reveals the average arrangement of the phases as a function of the position relative to the fracture initiation sites, by computing the probability of finding the hard phase around the fracture initiation sites. If at a certain relative position the probability of hard phase is higher than the volume fraction, $\langle \varphi^\mathrm{hard} \rangle$, having hard phase at this relative position promotes fracture initiation. Likewise, if the probability is lower than $\langle \varphi^\mathrm{hard} \rangle$, having soft phase at that relative position promotes damage.

The mathematical formalization is briefly repeated from \citep{DeGeus2015a}. It is first discussed based on a single volume element, but the results which are presented have additionally been average on all realizations in the ensemble. The distribution of phases is described using a so-called phase indicator, defined as follows
\begin{equation}
\mathcal{I} (i,j) =
\begin{cases}
	1 \quad &\text{if}~~(i,j) \in \text{hard}
	\\
	0 \quad &\text{if}~~(i,j) \in \text{soft}
\end{cases}
\end{equation}
whereby $(i,j)$ is the position of a grain within the volume element, in this case simply the row/column index in the regular grid. The damage weighted average phase, $\mathcal{I}_D$, at a certain distance $(\Delta i, \Delta j)$ from fracture initiation is then obtained by:
\begin{equation} \label{eq:hotspot}
  \mathcal{I}_D ( \Delta i , \Delta j ) = \frac{\sum_{i,j} D(i,j) \; \mathcal{I} ( i + \Delta i , j + \Delta j ) }{\sum_{i,j} D(i,j) }
\end{equation}
whereby $(i,j)$ loops over the grains in the volume element, taking the periodicity of the volume element into account\footnote{Note that \eqref{eq:hotspot} corresponds to a normalized discrete convolution between $\mathcal{I}$ and $D$. Therefore the evaluation for the special case of a square microstructure can be done using the discrete Fourier transform. }. The ensemble average $\langle \mathcal{I}_D \rangle$ trivially follows by looping over all volume elements in the ensemble. The quantitative interpretation is now as follows:
\begin{equation}
  \langle \varphi^\mathrm{hard} \rangle \; < \; \langle \mathcal{I}_D \rangle (\Delta i, \Delta j) \; \leq \; 1
\end{equation}
corresponds to an elevated probability of the hard phase at a relative position $(\Delta i, \Delta j)$ to the fracture initiation sites; and
\begin{equation}
  0 \; \leq \; \langle \mathcal{I}_D \rangle (\Delta i, \Delta j) \; < \; \langle \varphi^\mathrm{hard} \rangle
\end{equation}
to an elevated probability of the soft phase at that relative position.

\begin{figure}
  \centering
  \includegraphics[width=0.4\linewidth]{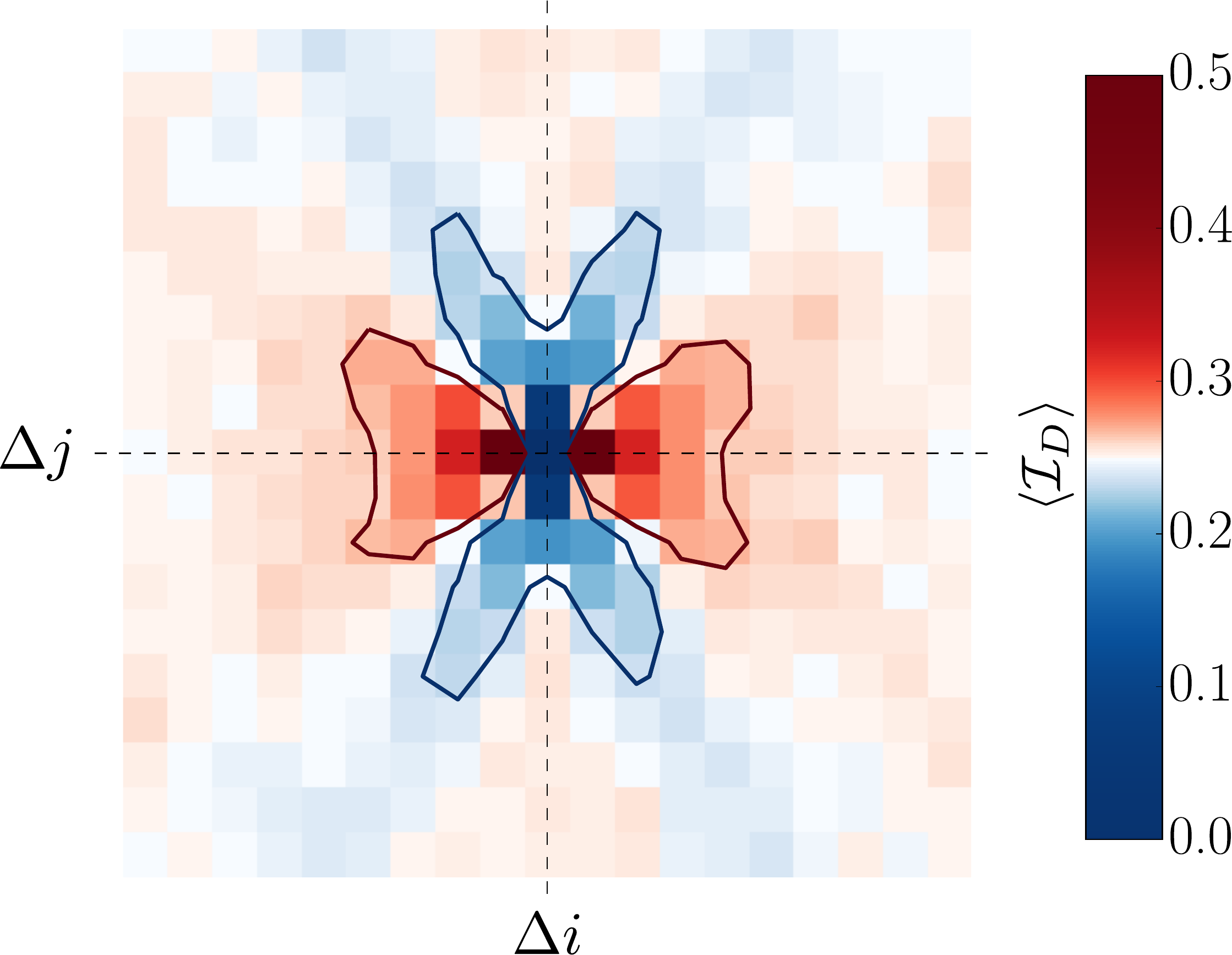}
  \caption{The ensemble averaged ``damage hot-spot'' $\langle \mathcal{I}_D \rangle$, at the final increment of deformation. The neutral color coincides with the average indicator, i.e.\ the hard phase volume fraction $\langle \varphi^\mathrm{hard} \rangle$, consequently red may be interpreted as a likelihood of the hard phase and blue as that of soft phase.}
  \label{fig:hotspot_ref}
\end{figure}

The result is included in Figure~\ref{fig:hotspot_ref}, taken at the final increment of deformation. The relative position $(\Delta i, \Delta j)$ to the initiation of fracture is indicated using dashed axes. The applied extension is in the horizontal direction and the compression is in the vertical direction. The colormap is defined such that the neutral color (white) corresponds to $\langle \varphi^\mathrm{hard} \rangle$; blue indicates an elevated probability of soft, and red an elevated probability of hard phase at that relative position to the fracture initiation sites. Iso-probability contours are used to highlight the main characteristics.

It is observed that -- by construction -- fracture initiates in the soft phase, indicated by the blue square at $(\Delta i, \Delta j) = (0,0)$. Directly to the right and left, in the direction of positive stretch, the probability of hard phase is close to one, while the probability of soft phase is one in the perpendicular direction. Further to the left and right an elevated probability of hard phase is found, while soft phase is found in bands at approximately $\pm 45$ degree angles.

These observations are the result of a combination of microstructural mechanisms: (i) a soft grain interrupting a band of hard phase along the tensile direction experiences high deformation, (ii) the phase boundary perpendicular to the tensile direction causes a hydrostatic tensile stress, and (iii) a soft band in the direction of shear, i.e.\ $\pm 45$ degrees, triggers high plastic deformation. The regions of hard phase in Figure~\ref{fig:hotspot_ref} are explained by the combination of (i) and (ii). In contrast, the orientation of the soft regions is the outcome of the competition between (ii) and (iii), resulting in an orientation between $0$ and $\pm 45$ degrees. More details of this hot-spot can be found in \citep{DeGeus2015a}.

\subsection{Simulations}

The simulations are performed using an optimized in-house code. For the domain size considered here, the time to compute the response of one volume element is approximately 15 minutes. Since the simulations can be done in parallel, the total time to compute the ensemble of volume elements is dependent on size and availability of the computing cluster.

\section{Influence of damage model}
\label{sec:damage}

\subsection{Johnson-Cook damage model}

The damage model as defined in equation~\eqref{eq:model:RT}, in the following referred to as the Rice \& Tracey model, captures the initiation of ductile fracture in a basic manner only. Most critically, the value of $D$ depends on the history only through the plastic -- deviatoric -- strain, and only on the current hydrostatic stress state. The influence of this assumption is assessed by using the more general history dependent Johnson-Cook model \citep{Johnson1985, Vajragupta2012}. In this case, the damage indicator is obtained by integrating the plastic strain rate $\dot{\varepsilon}_\mathrm{p}$ as a function of the stress state over the entire strain path. The damage indicator then reads
\begin{equation} \label{eq:model:JC}
	D = \int \frac{\dot{\varepsilon}_\mathrm{p}}{\varepsilon_\mathrm{c}(\eta)}~\mathrm{d}t
\end{equation}
where the critical strain $\varepsilon_\mathrm{c}$ is a function of the local stress triaxiality:
\begin{equation}
	\varepsilon_\mathrm{c} = A \exp \left( - B \eta \right) + \varepsilon_\mathrm{pc}
\end{equation}
wherein $A$, $B$, and the critical plastic strain $\varepsilon_\mathrm{pc}$ are material parameters. The stress triaxiality is defined in the usual way, as the ratio of the hydrostatic and equivalent (shear) stress:
\begin{equation}
\eta = \frac{\tau_\mathrm{m}}{\tau_\mathrm{eq}}
\end{equation}

The parameters of the Johnson-Cook model are based on the literature. For the class of materials considered in this paper the parameters are given by \citet{Vajragupta2012}:
\begin{equation} \label{eq:model:D_param}
A = 0.2 \qquad B = 1.7 \qquad \varepsilon_\mathrm{pc} = 0.1
\end{equation}
These parameters were obtained by fitting to response of the Gurson-Tvergaard-Needleman model for ductile fracture, for which the parameters had an experimental basis. Note again that the Rice \& Tracey model is normalized such that both models reach a value of $D = 1$ at the same applied uniaxial strain in a homogeneous soft material.

\subsection{Results and discussion}

\begin{figure}[tph]
  \centering
  \includegraphics[width=0.8\linewidth]{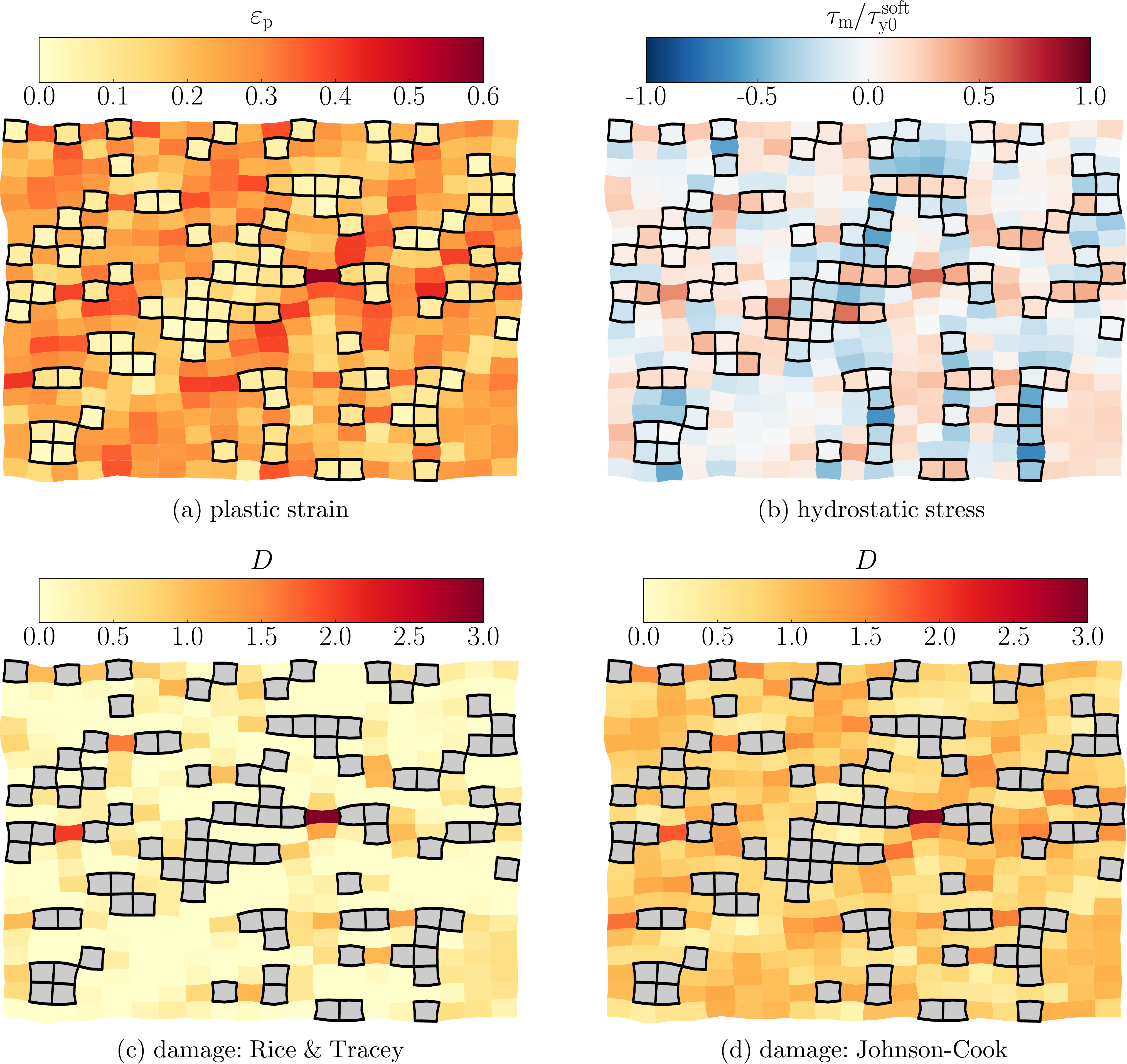}
  \caption{(a) The plastic strain $\varepsilon_\mathrm{p}$, (b) the hydrostatic stress $\tau_\mathrm{m}$, (c) the Rice \& Tracey damage indicator and (d) the Johnson-Cook damage indicator; for the microstructure from Figure~\ref{fig:typical}, at $\bar{\varepsilon}_\mathrm{d} = 0.2$. The hard cells are indicated using a black outline.}
  \label{fig:typical_damage}
\end{figure}

To illustrate the similarities and differences between the two models, the results for the two damage indicators in the microstructure of Figure~\ref{fig:typical} are shown in Figure~\ref{fig:typical_damage}(c--d); the plastic strain and hydrostatic stress in this microstructure are shown in Figure~\ref{fig:typical_damage}(a--b) at the same level of applied deformation. It is observed that the plastic strain, in Figure~\ref{fig:typical_damage}(a), is much higher in the soft phase than in the hard phase. The extremes occur in bands of connected soft phase under $\pm 45$ degree angles, observed most clearly in the band in the center of Figure~\ref{fig:typical_damage}(a). For the volumetric response it is observed that in the relatively large homogeneous region of soft phase, in the lower left part of Figure~\ref{fig:typical_damage}(b), the volumetric response coincides with that at the macroscopic level. Indeed, for the applied pure shear, the macroscopic volume is preserved and the hydrostatic stress vanishes in this region. In the other regions of Figure~\ref{fig:typical_damage}(b), the local mechanical incompatibility between the two phase causes a significant amount of local hydrostatic tension and compression.

The highest values of the damage indicator are observed only at a few locations in the microstructure which are identical for both damage indicators. The distinct characteristics of these hot-spot locations are clearly the same as in Figure~\ref{fig:hotspot_ref}: a soft grain neighbored to the left and right by hard grains and by soft grains on top and bottom. Also at larger distances the characteristics from Figure~\ref{fig:hotspot_ref} are easily identified. As observed from Figure~\ref{fig:typical_damage}(a--b), the hot-spots locally unite a high plastic strain with a positive hydrostatic stress, nearby a phase boundary.

For the Rice \& Tracey model in Figure~\ref{fig:typical_damage}(c) the damage is zero in many grains, in particular in regions with continuous soft phase. Although there is plastic strain, the hydrostatic stress is low in these regions (cf.\ Figure~\ref{fig:typical_damage}(a--b)). The difference for the Johnson-Cook model (Figure~\ref{fig:typical_damage}(d)) is that the value of the damage indicator is non-zero in almost every grain. This can be easily understood from equation~\eqref{eq:model:JC}, which shows that plastic straining contributes to the development of damage at any hydrostatic stress, even in hydrostatic compression (where $D$ increases, although at a low rate). In the microstructure, the highest values of both plastic strain and hydrostatic tension are caused by the mechanical incompatibility of the phases, resulting in the highest values of $D$. However, all soft grains experience some degree of plasticity, resulting in the lower, but non-zero, values of $D$.

\begin{figure}
  \centering
  \includegraphics[width=0.7\linewidth]{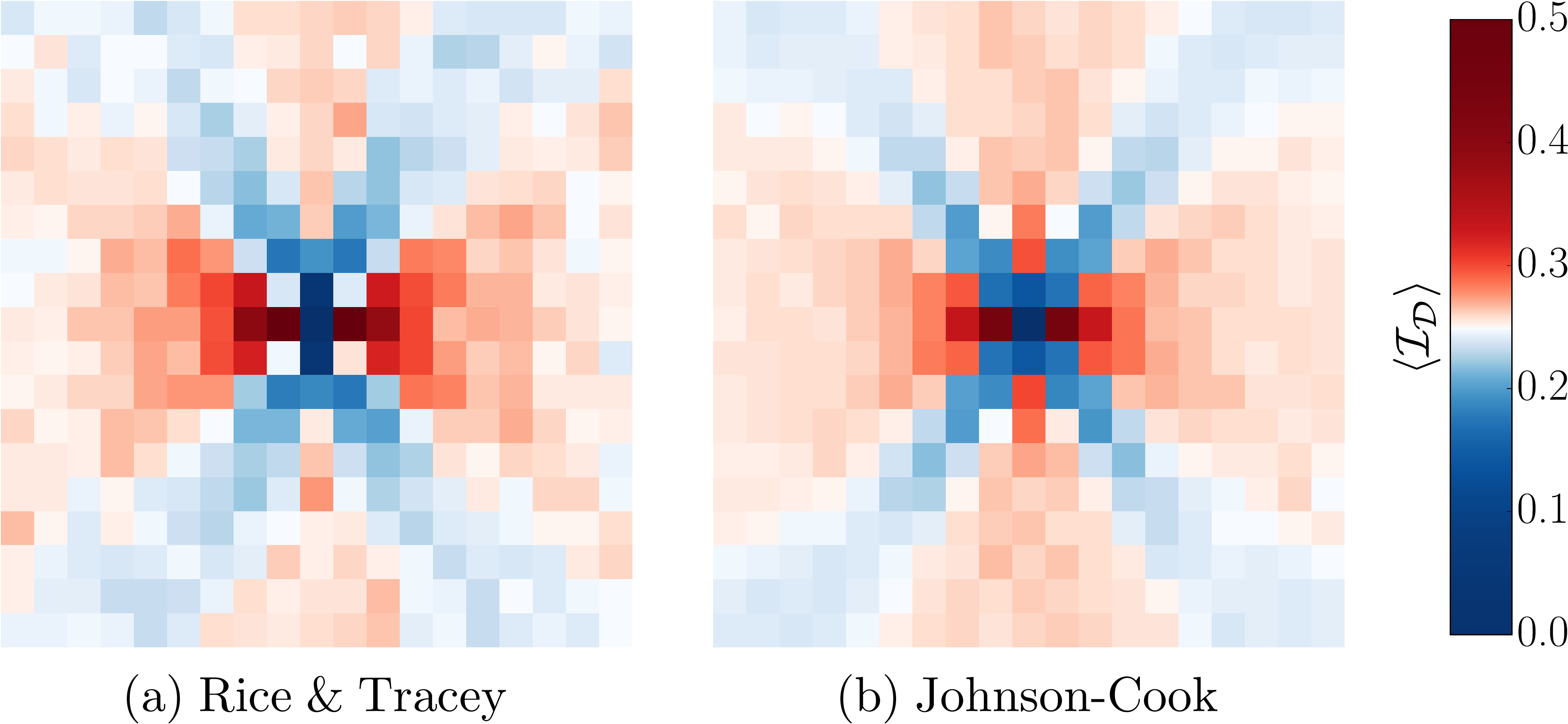}
  \caption{The ``damage hot-spot'' for the (a) Rice \& Tracey model and (b) the Johnson-Cook model. The neutral color of the colormap coincides with $\langle \varphi^\mathrm{hard} \rangle$.}
  \label{fig:hotspot_models}
\end{figure}

The fact that the Johnson-Cook damage indicator is also non-zero in grains where fracture initiation does not occur is inconvenient for our purpose. Particularly, the average microstructure around damage is in that case no longer governed by fracture initiation only. The damage indicator $D$ is therefore converted to a real fracture initiation indicator
\begin{equation}
  \mathcal{D} =
  \begin{cases}
  0 \quad & \mathrm{if}~~D  <   1 \\
  1 \quad & \mathrm{if}~~D \geq 1
  \end{cases}
\end{equation}
for both models. The ``damage hot-spot'' in equation~\ref{eq:hotspot} is calculated with this new definition. The resulting ``hot-spot'' is included in Figure~\ref{fig:hotspot_models} for (a) the Rice \& Tracey model and (b) the Johnson-Cook model. The first observation to be made is that the result for the Rice \& Tracey model using the fracture initiation indicator $\mathcal{D}$ coincides with the earlier result using the continuous $D$ (cf.\ Figure~\ref{fig:hotspot_models}(b) and Figure~\ref{fig:hotspot_ref}). The second observation is that the same qualitative pattern of the phase average distribution around the fracture initiation site is observed for the different damage models in Figure~\ref{fig:hotspot_models}. However, the orientation of the soft phase band is slightly different for the two damage models. The soft band for the Johnson-Cook model is closer to $\pm 45$ degrees. This is most obvious for an absolute (diagonal) distance $( | \Delta i | , | \Delta j |) = ( 1, 1 )$ for which soft phase is found for the Johnson-Cook model, but values close to $\langle \mathcal{I}_\mathcal{D} \rangle = \langle \varphi^\mathrm{hard} \rangle$ are found for the Rice \& Trace model. The orientation of the soft band is the outcome of a competition between (ii) a hydrostatic tensile stress caused by vertical phase boundaries and (iii) a high plastic strain caused by soft phase under $\pm 45$ degrees. As evidenced by Figure~\ref{fig:hotspot_models}, the different weighing of the plasticity and stress contributions in the two damage models leads to a slightly different outcome of this competition.

In conclusion the Rice \& Tracey model and the Johnson-Cook model lead to the same predicted location of fracture initiation. A small difference is observed in the average phase distribution around fracture initiation. In the following only the Johnson-Cook model is considered.

\section{Influence of the grain shape}
\label{sec:shape}

\subsection{Considered grain shapes}

\begin{figure}
  \centering
  \includegraphics[width=0.7\linewidth]{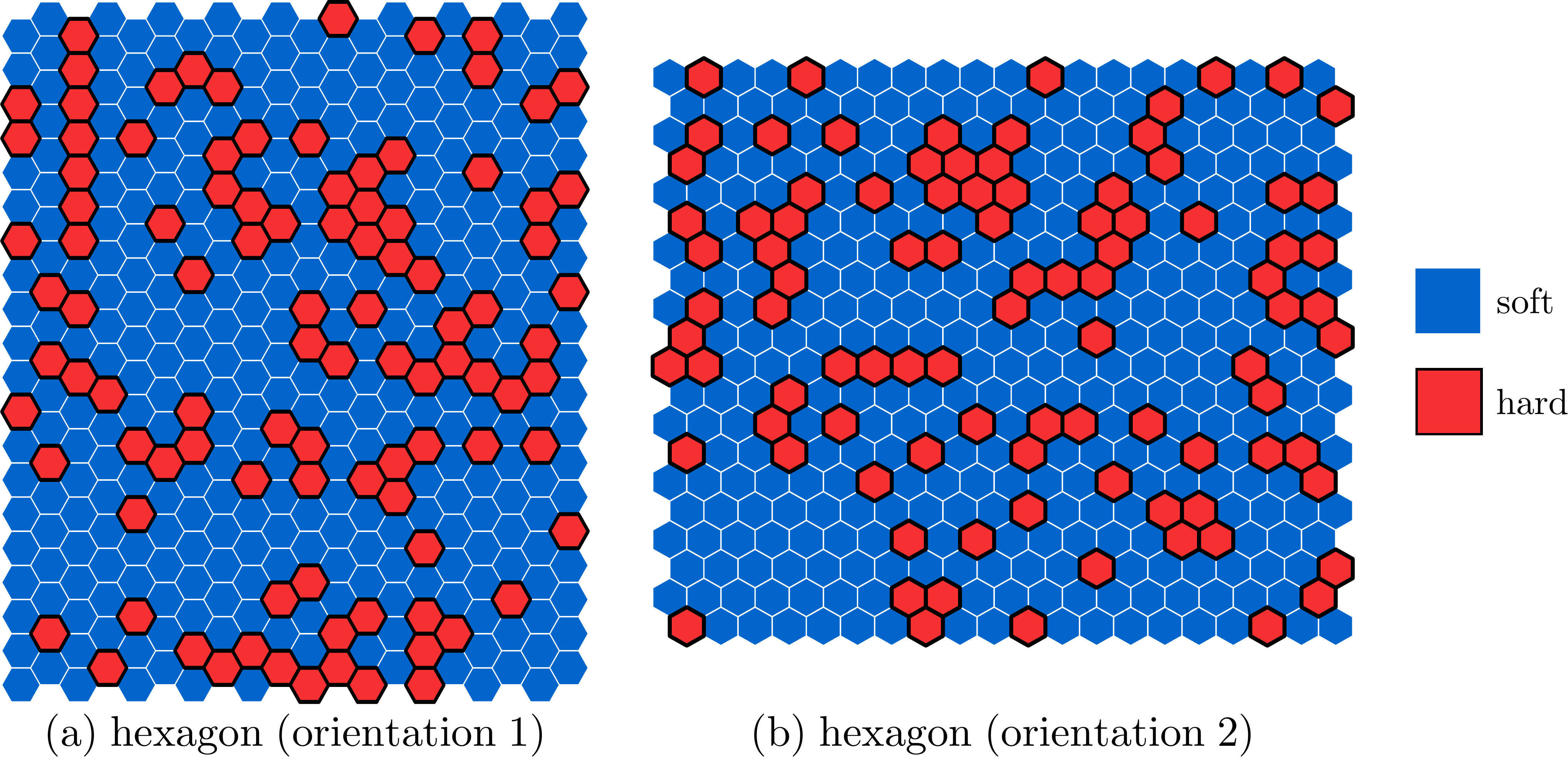}
  \caption{Typical volume elements using hexagonal grains in two orientations.}
  \label{fig:typical_hex}
\end{figure}

In order to assess the influence of the shape of the grains on the presented conclusions, the square grains are replaced by hexagonal grains with two different orientations (see Figure~\ref{fig:typical_hex}). A key difference between hexagonal grains and square grains is that the hexagonal grains have triple-junctions only (i.e.\ the corners are shared by three grains), whereas square grains reveal quadruple-junctions.

Furthermore, the square grains result in square volume elements. For this particular case the (plastic) shear bands may be influenced by the periodic boundary conditions \citep[e.g.][]{Coenen2012}. In contrast, equi-sized hexagonal grains result in a non-square volume element (cf.\ Figure~\ref{fig:typical_hex}(a--b)). Notice that the periodicity is slightly non-standard as one face of the grain in the corner is shared by more than one periodic repetition. In the case of square grains/elements this occurs only for a single point. See \citet{DeGeus2014} for details.

\subsection{Results}

\begin{figure}[tph]
  \centering
  \includegraphics[width=1.0\linewidth]{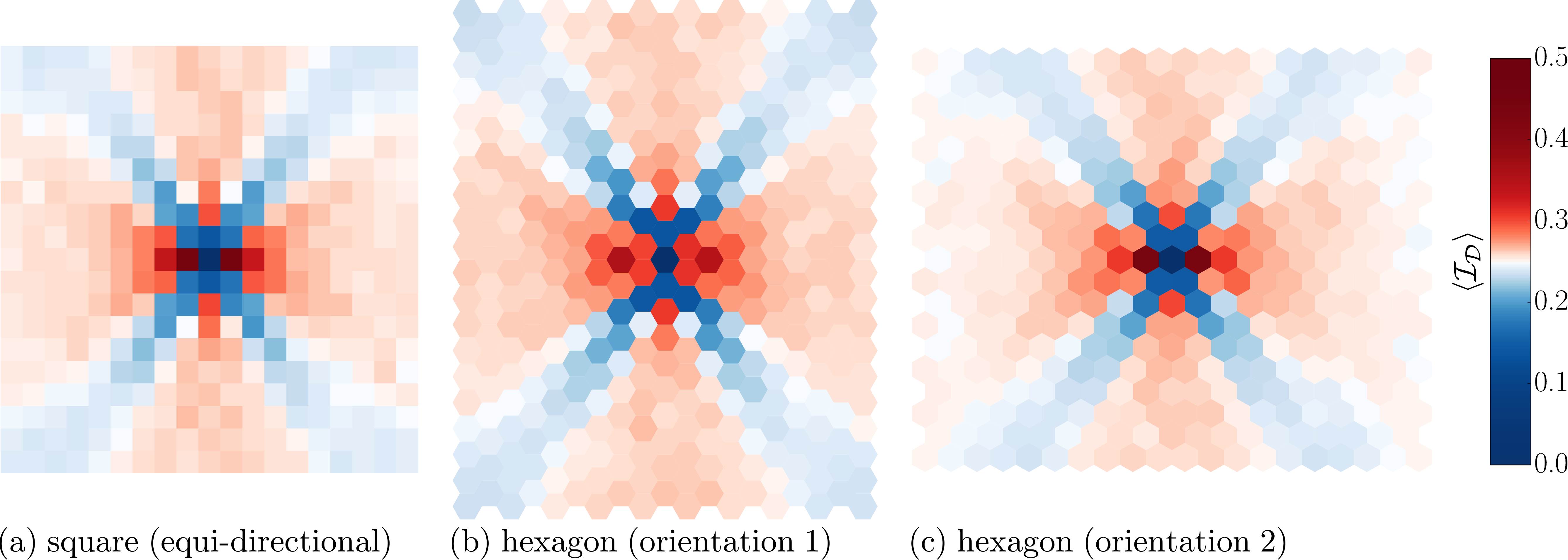}
  \caption{The average phase around the initiation of fracture at a deformation of $\bar{\varepsilon}_\mathrm{d} = 0.2$ for ensembles with different cell shapes. Blue may be interpreted as an elevated probability of soft phase, and red of hard phase. In each case shear is applied by extension in horizontal direction and compression in vertical direction.}
  \label{fig:hotspot}
\end{figure}

The resulting average phase distributions around the fracture initiation sites are depicted in Figure~\ref{fig:hotspot} for the three different grain shapes: square, horizontally and vertically oriented hexagons. The most important features are shared by all three configurations, and thus insensitive to the chosen grain shape and aspect ratio of the volume element. A small difference can be noticed between the horizontally oriented hexagons in Figure~\ref{fig:hotspot}(b) and the vertically orientated hexagons and the squares. Directly to the left and right from the central soft grain, a hard grain is found. However, the absolute probability (i.e.\ the value of $\mathcal{I}_\mathcal{D}$) in Figure~\ref{fig:hotspot}(b) is lower with respect to figures~\ref{fig:hotspot}(a) and (c). To understand this it is important to realize that the orientation of the soft band is the outcome of a competition between plasticity and hydrostatic tension. In the case of the horizontally oriented hexagons in Figure~\ref{fig:hotspot}(b) the microstructure is not able to capture the soft phase orientation that is `optimal' for the damage model like for the two other grain shapes. This is also observed in the values of the damage indicator (results not included), where for the configuration in Figure~\ref{fig:hotspot}(b) fracture initiation is predicted in 0.2\% fewer grains than for the other two grains shapes at the same applied strain $\bar{\varepsilon}_\mathrm{d} = 0.2$.

To summarize, the effect of the grain shape and the aspect ratio of the volume element seem to be minor. Fracture initiation is delayed slightly if the microstructure does not reflect the configuration that is critical for fracture initiation. Numerically this is observed when the volume element does not allow for this distribution.

\section{Influence of the stress state}
\label{sec:triax}

\subsection{Application of the hydrostatic stress}

So far, only a single deformation mode has been considered: pure shear. Since no volumetric deformation is applied in this mode, the overall hydrostatic stress vanishes. Note that this certainly does not imply a vanishing local hydrostatic stress -- even when averaged per grain -- see Figure~\ref{fig:typical_damage}(b). We next consider the effect of a macroscopic hydrostatic stress superimposed on the introduced shear stress. Since the damage development is dependent on the hydrostatic stress one may anticipate a significant effect of applying an overall hydrostatic stress on the damage distribution. To control the stress state, the macroscopic deformation gradient $\bar{\bm{F}}$ is decomposed in a purely volumetric part $\bar{\bm{F}}_\mathrm{v}$ and a purely isochoric part $\bar{\bm{F}}_\mathrm{d}$:
\begin{equation}
	\bar{\bm{F}} = \bar{\bm{F}}_\mathrm{v} \cdot \bar{\bm{F}}_\mathrm{d}
\end{equation}
herein $\bar{\bm{F}}_\mathrm{d}$ is taken identical to \eqref{eq:model:F} and $\bar{\bm{F}}_\mathrm{v}$ reads
\begin{equation}
	\bar{\bm{F}}_\mathrm{v} = \exp \big( \tfrac{1}{3} \varepsilon_\mathrm{v} \big) \, \bm{I}
\end{equation}
with $\bm{I}$ the identity tensor and $\varepsilon_\mathrm{v}$ the volumetric logarithmic strain. The latter is evolved such that at any time during the simulation the resulting overall triaxiality $\bar{\eta}$ is constant and given by
\begin{equation}
	\bar{\eta} = \frac{\bar{\tau}_\mathrm{m}}{\bar{\tau}_\mathrm{eq}}
\end{equation}
with $\bar{\tau}_\mathrm{m}$ and $\bar{\tau}_\mathrm{eq}$ the macroscopic hydrostatic and von Mises equivalent Kirchhoff stress, respectively. A trivial case is $\bar{\eta} = 0$ for which $\bar{\tau}_\mathrm{m} = 0$ and thus $\varepsilon_\mathrm{v} = 0$, as used so far in the previous sections.

The equivalent stress is obtained through homogenization. To this end the macroscopic Kirchhoff stress is decomposed in a volumetric part and a deviatoric part:
\begin{equation} \label{eq:stress_split}
	\bar{\bm{\tau}} = \bar{\tau}_\mathrm{m} \bm{I} + \bar{\bm{\tau}}^\mathrm{d}
\end{equation}
where the deviatoric stress is a function of the microstructure (cf.\ Figure~\ref{fig:stress-strain}). The hydrostatic stress can be explicitly expressed in terms of the volumetric elastic strain by virtue of the assumption that the microstructure is elastically homogeneous (and that the plastic response is purely deviatoric). Following (\ref{eq:model:stress}--\ref{eq:model:stiff}) the macroscopic hydrostatic stress
\begin{equation}
	\bar{\tau}_\mathrm{m} = K \varepsilon_\mathrm{v}
\end{equation}
and therefore the macroscopic triaxiality
\begin{equation} \label{eq:triax:mu-taueq}
	\bar{\eta} = \frac{ K \varepsilon_\mathrm{v} }{ \bar{\tau}_\mathrm{eq} }
\end{equation}
Note that $\bar{\tau}_\mathrm{eq}$ is a function of the microstructure, and thus a result of the simulation. To avoid enforcing \eqref{eq:triax:mu-taueq} iteratively during the simulation, the equivalent stress at the previous increment is used, i.e.\
\begin{equation}
	\varepsilon_\mathrm{v}^{(t + \Delta t)} = \frac{ \bar{\eta}\, \bar{\tau}_\mathrm{eq}^{(t)} }{ K }
\end{equation}
This explicit integration results in a small underestimation of $\bar{\eta}$ (less than 1\%).

In the results section, the macroscopic triaxiality is varied in the range
\begin{equation}
	- 0.2 < \bar{\eta} < 1.0
\end{equation}

\subsection{Results}

\begin{figure}[tph]
  \centering
  \includegraphics[width=0.7\linewidth]{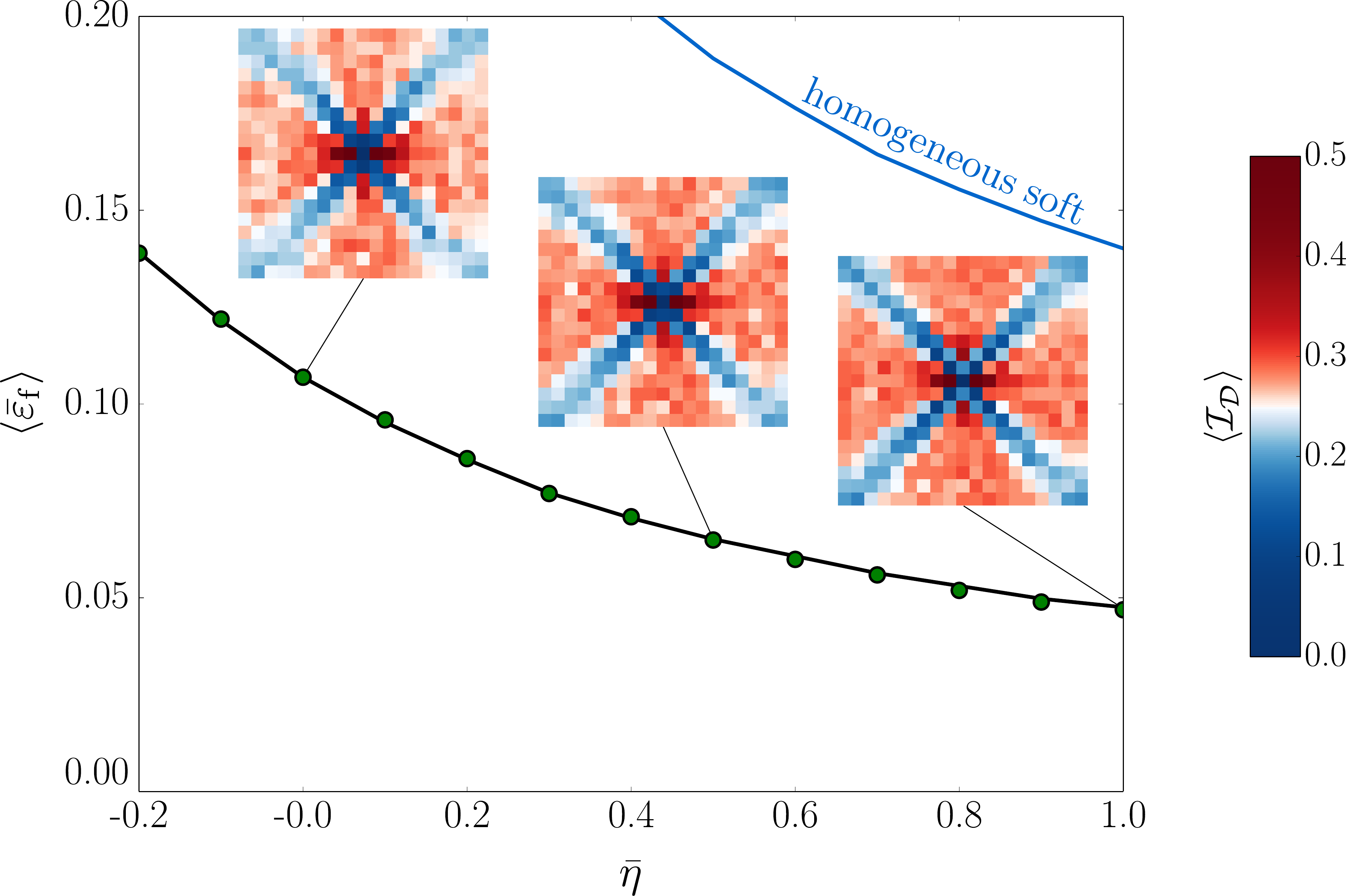}
  \caption{The macroscopic fracture initiation strain $\langle \bar{\varepsilon}_\mathrm{f} \rangle$ as a function of the applied macroscopic stress triaxiality $\bar{\eta}$. The limit for which the microstructure is homogeneously soft is included as a blue curve. For the macroscopic triaxialities $\bar{\eta} = 0$, $0.5$, and $1$, the average microstructural morphology around the individual fracture initiation sites is shown, the colorbar corresponds to these three plots.}
  \label{fig:fracture}
\end{figure}

The results are analyzed in terms of macroscopic fracture initiation. A simple criterion is introduced whereby macroscopic fracture is expected to occur, e.g.\ when fracture has initiated in 1\% of the grains in the ensemble. The predicted macroscopic equivalent strain at which fracture initiates, $\langle \bar{\varepsilon}_\mathrm{f} \rangle$, is plotted in Figure~\ref{fig:fracture} as a function of the applied triaxiality $\bar{\eta}$ using a black line. The limit for the homogeneous soft phase is included as a blue line. It is observed that $\langle \bar{\varepsilon}_\mathrm{f} \rangle$ decreases with increasing applied $\bar{\eta}$, as suggested by the dependence of the damage indicator on the local triaxiality (see equation~\eqref{eq:model:JC}). The macroscopic fracture initiates at a significantly lower strain for the two-phase material than for the homogeneous soft phase, whereby a slightly different dependence on $\bar{\eta}$ is observed for the two-phase microstructure than for the homogeneous soft phase.

Next, the individual fracture initiation sites are studied in more detail. The average phase distribution around the microstructural fracture initiation sites is included for three representative macroscopic triaxialities, $\bar{\eta} = 0$, $0.5$, and $1$\footnote{Notice that the earlier results (Figures~\ref{fig:hotspot_models} and \ref{fig:hotspot}) are at a different applied strain level.}. Qualitatively, each of these triaxialities share the same pattern, with hard phase in $x$-direction and soft phase under angles close to $\pm 45$ degrees. A minor change in orientation of the soft bands is however observed. For higher values of $\bar{\eta}$ the presence of vertical phase boundaries is less important as the hydrostatic tensile state is partly caused by the macroscopically hydrostatic tension. This results in a soft phase band which more closely approximates $\pm 45$ degree angles with increasing triaxiality.

\subsection{Reduction of the computational costs}

\begin{figure}
\centering
\includegraphics[width=1.0\linewidth]{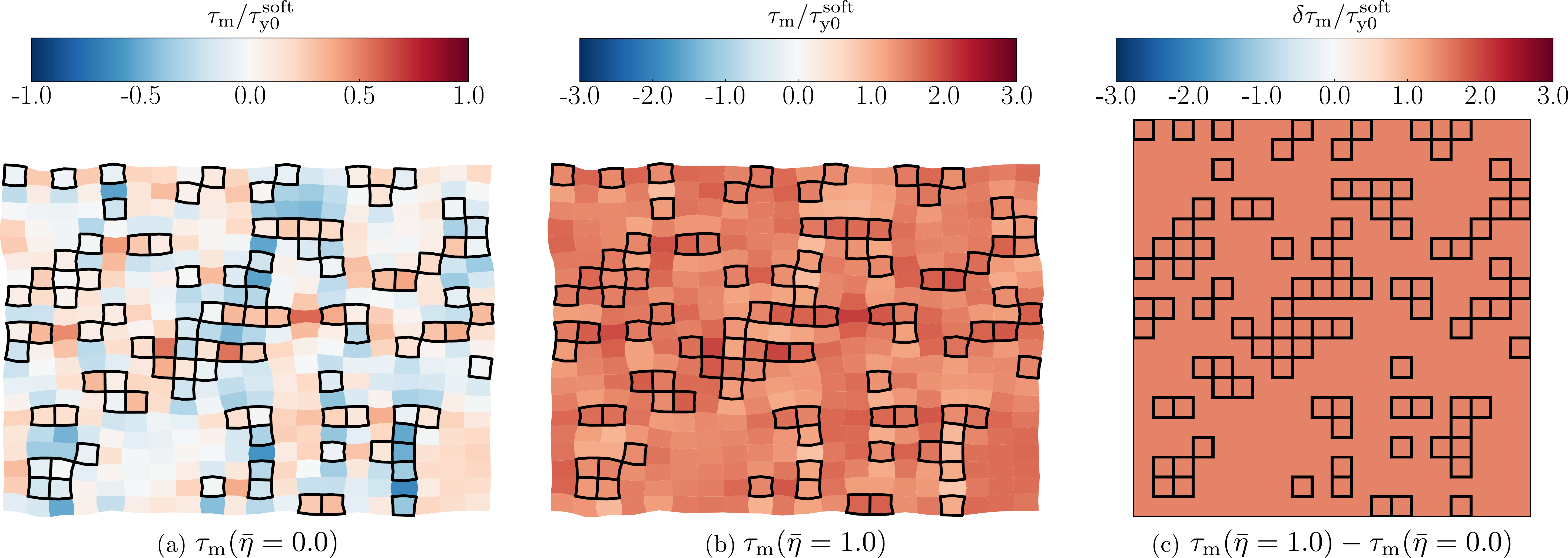}
\caption{The local hydrostatic stress at (a) $\bar{\eta} = 0$ and (b) $\bar{\eta} = 1$, at $\bar{\varepsilon}_\mathrm{d}$. The difference: (a) subtracted from (b) is show in (c). Notice the different scales bars used.}
\label{fig:triax_taum}
\end{figure}

The fact that the volumetric response of the two phases is elastic and that their elastic properties are identical implies that the increase of hydrostatic stress due to the applied $\bar{\eta}$ is homogeneous. This is illustrated using the microstructure given in Figure~\ref{fig:typical}. The local hydrostatic stress is shown in Figure~\ref{fig:triax_taum}(a--b) for $\bar{\eta} = 0$ and $1$ respectively. The local difference -- i.e.\ Figure~\ref{fig:triax_taum}(a) subtracted from Figure~\ref{fig:triax_taum}(b) -- is shown in Figure~\ref{fig:triax_taum}(c). Indeed, the increase in hydrostatic stress is homogeneous.

This suggest that the response for different applied triaxialities can be closely approximated by simply superimposing a uniform hydrostatic stress on the stress field of a single simulation at $\bar{\eta} = 0$ for microstructures with a homogeneous volumetric, but inhomogeneous deviatoric properties. To this end the local hydrostatic stress $\tau_\mathrm{m}$ is split into a local contribution $\tilde{\tau}_\mathrm{m}$ and a global contribution $\bar{\tau}_\mathrm{m}$:
\begin{equation}
  \tau_\mathrm{m}( \vec{x}, \bar{\bm{\varepsilon}} ) = \tilde{\tau}_\mathrm{m}( \vec{x}, \bar{\bm{\varepsilon}} ) + \bar{\tau}_\mathrm{m}( \bar{\bm{\varepsilon}} )
\end{equation}
wherein the dependence on the position $\vec{x}$ and the applied deformation $\bar{\bm{\varepsilon}}$ have been explicitly specified. The local hydrostatic stress $\tilde{\tau}_\mathrm{m}$ is obtained from equilibrium for a specific macroscopic stress state in which the hydrostatic component vanishes, for example using \eqref{eq:model:F}. The average hydrostatic stress $\bar{\tau}_\mathrm{m}$ is then determined from the applied triaxiality $\bar{\eta}$ as follows
\begin{equation}
  \bar{\tau}_\mathrm{m} ( \bar{\bm{\varepsilon}} ) = \bar{\eta} (\bar{\bm{\varepsilon}} )\, \bar{\tau}_\mathrm{eq} (\bar{\bm{\varepsilon}})
\end{equation}
where typically $\bar{\eta}$ is kept constant throughout the deformation history. This method results in stress state which is in equilibrium, however the predicted strains are not fully compatible with these stresses.

In this simplified method, a single simulation is performed (for each volume element in the ensemble). Based on this one simulation the full stress tensor is calculated at each time increment, for all the different triaxialities. The integration of the damage indicator over the history is thus done for the different triaxialities individually, but is based on a single finite element calculation. This thus greatly reduces the computational cost.

The predicted macroscopic fracture strain $\langle \varepsilon_\mathrm{f} \rangle$ using a single simulation (of the ensemble) in pure shear is included in Figure~\ref{fig:fracture} using green markers. As observed, the difference due to the (must faster) approximation are extremely small (less than 1\%). The largest source in inaccuracy is the discretization in time (i.e.\ the size of the deformation increments). The predicted macroscopic logarithmic strain $\bar{\varepsilon}$ is furthermore a bit too small as it does not account for the elastic volumetric contribution. However, these strains are small in comparison to the deviatoric strains as the bulk modulus $K$ is much larger than the hardening moduli $H$ of the two phases.

\section{Discussion and concluding remarks}
\label{sec:discussion}

Earlier work has identified the role of the local spatial distribution of phases on the onset of ductile fracture in a two-phase microstructure using an idealized model subjected to a single stress state \citep{DeGeus2015a}. It was observed that elongated regions of hard phase in the tensile direction interrupted by soft bands under $\pm 45$ degree angles -- in the direction of shear -- are the most likely locations for the fracture to initiate. The exact orientation of the band of soft phase with respect to the tensile axis is determined by the competition of the hydrostatic tensile stress (caused by phase boundaries perpendicular to the tensile axis) and plastic strain (caused by soft bands under exactly $\pm 45$ degrees).

Since the conclusions in \citep{DeGeus2015a} may be sensitive to modeling assumptions, this work critically assesses three major assumptions made in \citep{DeGeus2015a}, resulting in the following conclusions:
\begin{enumerate}
  \item The simple damage indicator was replaced by the more realistic Johnson-Cook model, which includes the influence of the stress state throughout the deformation history. The different damage models qualitatively yield the same spatial phase distribution around the fracture initiation site, but have a slightly different orientation of the band of soft phase.
  \item Changing the grain shape from square to hexagonal has a limited effect on the predicted fracture initiation. The only significant influence is the ability to capture a critical phase distribution around the fracture initiation site as predicted by the damage indicator. If the grain shape is unfavorable in relation to the critical configuration fracture initiation is slightly delayed. The effect of different volume element aspect ratios is small, suggesting that the results are not dominated by the periodicity assumption. Note that at the local level of the sub-grain, the grain shape may be more important \citep[observed for example in][]{Williams2012, Llorca1991}. Here, we only show that on the level of aggregates of grains these difference may not be essential.
  \item Different stress states, in particular higher stress triaxialities, yield a different overall ductility but similar critical phase distributions around the fracture initiation site. The ``ideal'' orientation of the band of soft phase with respect to the tensile axis approaches $\pm 45$ degrees for increasing triaxiality, as the tensile hydrostatic stress no longer depends on the phase boundaries perpendicular to the tensile axis. A novel simplified approximation was proposed to analyze the influence of the applied stress triaxiality using a single (set of) deformation controlled simulations. In this way, numerical issues related to prescribed forces \citep[e.g.][]{Kuna1996} are avoided.
\end{enumerate}

In the present study a structured two-dimensional microstructure was used, which enabled a systematic analysis. However, the methods developed lend themselves well to extend the analysis to a more realistic setting. First, the analysis may be extended to three dimensions a straightforward fashion, except for the significantly increased computational cost. A preliminary analysis, using a coarse numerical discretization, in \citep{DeGeus2015a} has shown that for the same applied deformation the damage response is not strongly affected by limiting the analysis a two-dimensional microstructure. However, a more systematic analysis that includes different deformation states is needed. Second, the analysis can be applied to a microstructural description characterized by a more realistic distribution of grain sizes and shapes, whereby the computation of the average phase distribution around fracture initiation should carefully account for the presence of more than one length scale. However, to model a set of microstructures that exactly matches such a distribution is far from trivial and not yet fully developed \citep[e.g.][]{Kumar2006, Schroder2010}. Finally, one could image the hot-spot analysis directly to microscopic images that comprise both the microstructure and the fracture sites.

Several other assumptions need attention in future work. The hard phase was assumed not to fracture. However, depending on the considered material and stress state this may be a restrictive assumption \citep{Ahmad2000, Avramovic-Cingara2009, Avramovic-Cingara2009a, Uthaisangsuk2008}. This study was limited to the initiation of fracture. Different mechanisms may be controlling for the propagation of (macroscopic) fracture.

\section*{Acknowledgments}

This research was carried out under project number M22.2.11424 in the framework of the research program of the Materials innovation institute M2i (\href{http://www.m2i.nl}{www.m2i.nl}).

The authors express their gratitude to the organizers of the 20th European Conference on Fracture (ECF20) and the European Structural Integrity Society (ESIS) for their efforts in the publication of this special issue.


\scriptsize
\bibliography{library}

\end{document}